\documentstyle[aps]{revtex}
\begin{document}
\preprint{ {\bf /} {\bf /} }
\draft
\title{Symbolic Dynamics of the Diamagnetic Kepler Problem 
Without Involving Bounces}
\author{ Zuo-Bing Wu$^{1}$ and  Wei-Mou Zheng$^{2}$} 
\address{$^1$Department of Physics, Peking University,
Beijing 100871, China\\
$^2$Institute of Theoretical Physics, Academia Sinica,
Beijing 100080, China}
\date{\today}
\maketitle

\begin{abstract}
Without involving bounce events, a Poincar\'e section associated with the 
axes is found to give a map on the annulus for the diamagnetic Kepler 
problem. Symbolic dynamics is then established based on the lift of the 
annulus map. The correspondence between the coding derived from this axis 
Poincar\'e section is compared with the coding based on bounces. Symmetry 
is used to reduce the symbolic dynamics. By means of symbolic dynamics the 
admissibility of periodic orbits is analyzed, and the symmetry of 
orbits discussed.
\end{abstract}

\section{Introduction}

The diamagnetic Kepler problem (DKP), or a hydrogen atom moving 
in a uniform magnetic field, is a simple physical system for 
studying the correspondence relation between classical chaos
and quantum behavior\cite{FW}. In the semiclassical quantization 
of a classically chaotic system, quantum properties are related to 
periodic orbits or closed orbits\cite{DD,WH,THM}. An 
effective method to locate orbits, and then encode and classify them is 
essential to the study. For the $n$-disk scattering system\cite{Hansen1} 
and stadium billiard\cite{HC,Zheng1}, symbolic dynamics has been 
established by taking a Poincar\'e section on the boundary of reflection.  
The symbolic encoding of the orbits and symbolic dynamics developed 
for the four-disk system have been extended to the smooth Hamiltonian 
system for the DKP\cite{EW,Hansen2,HG}. However, bounces are not
always well defined for a soft potential. For the hyperbolic potential 
system, symbolic dynamics is established by introducing an appropriate 
Poincar\'e section without referring to bounce  events\cite{Zheng2}.
Here we use this method to construct symbolic dynamics for the DKP. 
The paper is organized as follows. In Sec. \ref{sec:poin_sect}, we describe 
the method to construct a Poincar\'e section with the Birkhoff coordinates.
In Sec. \ref{sec:symb_dyna} we discuss the symbolic encoding of orbits
and ordering rules of symbolic sequences. We examine the correspondence 
between this coding and the coding based on bounces for the touching 
four-disk billiard in Sec.\ \ref{sec:4disk}. In Sec. \ref{sec:encd_class}, 
we classify orbits according to their codes and symmetries. Finally, in Sec. 
\ref{sec:sum}, we give a summary and conclusion.

\section{The Poincar\'e section}
\label{sec:poin_sect}

The classical dynamics of a hydrogen atom in a uniform magnetic field $B$ 
along the $z$-axis is described by the Hamiltonian:
\begin{equation}
H=\frac{p^2}{2m_e}-\frac{e^2}{r}+\omega l_z+\frac{1}{2}m_e \omega^2 (x^2+y^2)
\label{1}
\end{equation}
where $l_z$ is the $z$ component of the angular momentum, and $\omega=
\frac{eB}{2mc}$ the cyclotron frequency. 
In cylindrical coordinates ($\rho$,$z$,$\phi$) and atomic units, at the 
angular momentum $l_z=0$, the Hamiltonian can be
rewritten as
\begin{equation}
H=\frac{p_{\rho}^2}{2}+\frac{p_z^2}{2}-\frac{{\rho}^2}{8}-\frac{1}
{\sqrt{{\rho}^2+z^2}},
\label{2}
\end{equation}
which is an even function of the coordinates and momenta. Its 
symmetry group consists of the identity $e$ and the reflection $\sigma_{\rho}$. 
Introducing the semi-parabolic coordinates
\begin{eqnarray}
\mu^2 =&&\sqrt{\rho^2+z^2}-z,\nonumber\\
\nu^2 =&&\sqrt{\rho^2+z^2}+z,\\
d \tau=&&\frac{1}{2} dt/ \sqrt{\rho^2+z^2},\nonumber\\
\end{eqnarray}
we may convert the Hamiltonian to
\begin{equation}
h=\frac{p_\mu^2}{2}+\frac{p_\nu^2}{2}-\epsilon(\mu^2+\nu^2)+\frac{1}{8}\mu^2\nu^2(\mu^2+\nu^2)
\equiv 2,
\label{3}
\end{equation}
where $\epsilon=E (2\omega )^{-2/3}$ is the scaled energy and
$h$ is the fixed ``pseudo energy''. In Fig.\ 1, we show an orbit at 
$\epsilon=0$. Irregular orbits at $\epsilon=0$ displayed in the configuration 
space($z$, $\rho$)\cite{DD} or in the Poincar\'e section ($\nu$, 
$p_{\nu}$)\cite{FW} show the chaotic behavior of the dynamics. 

The transformed Hamiltonian $h$ has a higher symmetry than $H$. Its symmetry 
group consists of the identity $e$, two reflections $\sigma_{\mu}$, 
$\sigma_{\nu}$ across the $\mu$, $\nu$ axes, two diagonal reflections 
$\sigma_{13}$, $\sigma_{24}$, and three rotations $C_4$, $C_2$ and $C_4^3$ 
by $\pi/2$, $\pi$ and $3\pi/2$ around the center, respectively\cite{CE}.

Our Poincar\'e section is chosen as follows. Imagine that the $\mu$ and $\nu$ 
axes are both of a finite width and length. A counterclockwise contour is 
taken along the perimeter of the area forming by the two crossing imaginary 
rectangles. The Poincar\'e section is then obtained by recording the 
position and the tangent component of the momentum along the contour, i.e., 
the Birkhoff canonical coordinates \cite{Birkhoff} at intersecting points 
with the contour where an orbit enter the inside of the contour. The length 
of the contour is infinite. It is more convenient to transform the contour 
to one with a finite length. For example, in the first quadrant, the 
transformations $s=-\mu /(1+\mu )$ along the positive $\mu$ axis and 
$s=\nu /(1+\nu )$ along the positive $\nu$ axis convert the segment of the 
original contour in the first quadrant into an interval of length 2 
parametrized with $s\in [-1,1)$. For our purpose here, the variable 
corresponding to the momentum may be take as $v=-p_\mu /p$ at the positive 
$\mu$ axis, and $v=p_\nu /p$ at the positive $\nu$ axis, where $p=\sqrt{
p_{\mu}^2+p_{\nu}^2}$. In this way we may parametrize the whole contour 
with $s\in [-1,7)$, and define corresponding $v$. The rotational symmetry 
under $C_4$, $C_2$ and $C_4^3$ in the original configurational space 
becomes the translational symmetry of shifting $s$ by a multiple of 2 in 
the $s$-$v$ plane. The dynamics on the Poincar\'e surface is then 
represented by a map on the annulus $s\in [-1,7)$ and $v\in [-1,1]$. 

\section{Symbolic Dynamics Based on Lifting}
\label{sec:symb_dyna}
For the map on an annulus describing the dynamics on the Poincar\'e section,
it is useful to consider its lift. In the lifted space on which the lift map
is defined the area $s\in [-1,7)$ and $v\in [-1,1]$ consists of a fundamental
domain (FD), which is uniformly divided into 8 strips according to $s$.
The image of the FD in the lifted space is partly shown in 
Fig.~\ref{fig:fd}(a),
where the 8 strips of the FD are marked with the numbers 0 to 7, and four
strips right of the FD marked with $+0$ to $+3$. In the figure we draw only
the images of the strips 1 and 2. Two zones marked with $1$ and $1'$ of
the strip 1 are mapped to the zone 1 of strip $+3$ and the zone $1'$ of
strip $+2$, respectively. The notation representing the image of strip 2
is analogous. The rotational symmetry of the Hamiltonian 
corresponds to the translational symmetry in the lifted space. For example,
the image of strip 0 is in the strips 6 and 7, and can be obtained from
the image of strip 2 (in the strips $+0$ and $+1$) by shifting to the
left by 2 strips. Similarly, we show part of the preimage of the FD in
Fig.~\ref{fig:fd}(b). Since on the annulus forward and backward 
maps rotate in opposite directions, strips are on the right of 
their preimages, in contrast with images.

For the full FD, the rotation number of orbits is between 0 and 2. Taking
the rotational symmetry into account, we may consider only the reduced
domain (RD) consisting of strips 1 and 2. When the RD is regarded as the 
annulus, from Fig.~1 the integral part of the rotation numbers is 3 and 4
for zone 2, which is the joint zone of $1'$ and $2'$, but 4 and 5 for 
zone 1. As shown in Fig.\ \ref{fig:rd}(a), the RD may be correspondingly 
divided into three regions according to the rotation 
number. The two lines of demarcation are marked by $B_0\bullet$ and $B_2
\bullet$. In the figure we have also drawn some stable and unstable 
manifolds. (The stable manifolds have a segment roughly parallel to lines 
$B_0\bullet$ and $B_2\bullet$, and the unstable manifolds are a mirror 
image of the stable manifolds with respect to the line $v=0$.) In the RD, 
tangencies between stable and unstable manifolds can be easily seen. Among 
them, the prominent ones are in the top-left and bottom-right regions of 
the three. At a given point the stable and unstable directions, i.e. tangent
directions of manifolds, can be determined with the procedure suggested by
Greene \cite{Greene}, and then tangent points found. Two lines connecting such
tangent points, which are marked by $C_0\bullet$ and $C_2\bullet$, give
a further partition of the RD. The final partition of the RD into five
regions is shown in Fig.\ \ref{fig:rd}(a), where these regions are marked by 
$L_0\bullet$, $R_0\bullet$, $R_1\bullet$, $R_2\bullet$ and $L_2\bullet$. In the
figure the two boundaries of the RD are marked by $D_0\bullet$ and
$D_2\bullet$. This partition is the partition according to the preimage. Its 
preimage gives the partition according to the present, as shown in 
Fig.~\ref{fig:rd}(b), 
where the five regions are marked with $\bullet L_0$, $\bullet R_0$, 
$\bullet R_1$, $\bullet R_2$, and $\bullet L_2$. By means of this 
partition, or equivalently the partition of Fig.~\ref{fig:rd}(a) according to 
the preimage, we may code an orbit with a doubly infinite sequence
$$\cdots s_{-1}\bullet s_0 s_1\cdots ,$$
where $\bullet$ indicates the present. We may refer to this coding as the 
axis coding, and the coding based on bounces in the literature 
the bounce coding.

The ordering is essential to establishing symbolic dynamics. The natural 
order is well defined in the lifted space. The 5-piece preimage of the RD 
arranged from left to right in the lifted space consist of $L_0\bullet$, 
$R_0\bullet$, $R_1\bullet$, $R_2\bullet$ and $L_2\bullet$, as can be seen 
from Fig.~\ref{fig:fd}(b). Thus, according to the natural order we have
\begin{equation}
L_0\bullet < R_0\bullet < R_1\bullet <R_2\bullet <L_2\bullet.  
\label{bord}
\end{equation}
From the preimage of the RD shown in Fig.~\ref{fig:fd}(b) we see an opposite 
arrangement 
of the preimage pieces in the lifted space. A more precise description of 
ordering is given by the ordering of unstable manifolds on a transversal 
stable foliation. It is numerically verified that under the backward map 
the ordering is reversed in the regions $R_0\bullet$, $R_1\bullet$ and 
$R_2\bullet$, but preserved in $L_0\bullet$ and $L_2\bullet$. We may define 
the parity of a finite string by the oddness of the total number of the 
letters $R_0$, $R_1$ and $R_2$ contained in the string. Any odd leading 
string in backward sequences will then reverse the ordering (\ref{bord}) of 
backward sequences. 

Similarly, from the partition shown in Fig.~\ref{fig:rd}(b) according to 
the present, we have the ordering 
\begin{equation}
\bullet L_0< \bullet R_0< \bullet R_1< \bullet R_2< \bullet L_2.
\label{ford}
\end{equation}
An odd leading string also reverses this ordering for forward sequences. 
 
Based on the ordering rules, metric representations for both forward and
backward sequences may be introduced to construct the symbolic plane\cite{CGP}.
Every forward sequence $\bullet s_0s_1\cdots s_n\cdots$ may correspond to a 
number $\alpha$, represented in base 5, between 0 and 1 as follows. We define 
the correspondence of symbol $s_i$ to number $\mu_i\in\{0,1,2,3,4\}$ as $L_0 
\to 0$, $R_0 \to 1$, $R_1 \to 2$, $R_2 \to 3$ and $L_2 \to 4$ if the 
leading string $s_0s_1\cdots s_{i-1}$ is even, and as $\{L_0,R_0,R_1,R_2,
L_2\} \rightarrow \{4,3,2,1,0\}$ otherwise. Finally, we define
 \begin{equation}
 \alpha  = \sum_{i=0}^\infty \mu_i 5^{-(i+1)}.
 \label{eq4} 
 \end{equation}
In this way forward sequences are ordered according to their $\alpha$-values. 
Similarly, we assign to a backward sequence $\cdots s_{-m} \cdots s_{-2} 
s_{-1}\bullet$ the number $\beta$ defined by
 \begin{equation}
 \beta = \sum_{j=1}^\infty \nu_j 5^{-j},
 \label{eq5}
 \end{equation}
where $\nu_j\in\{0,1,2,3,4\}$ is determined by $s_{-j}$ and the oddness of 
the leading string $s_{-j+1}\cdots s_{-2}s_{-1}$. In the symbolic plane 
every orbit point corresponds to a point ($\alpha$, $\beta$) in the unit 
square, where $\alpha$ and $\beta$ are associated with the forward and 
backward sequences of the orbit point, respectively. We show the symbolic 
plane for $\epsilon=0$ in Fig.\ \ref{fig:rdsp}(a), where approximately 68000 
points of 
several real orbits in the RD are drawn. The corresponding pruning front for 
the partition lines $\bullet C_0$, $\bullet C_2$ and the boundary lines 
$\bullet D_0$ and $\bullet D_2$ is shown in Fig.\ \ref{fig:rdsp}(b).

So far we have not considered the reflectional symmetry. In the lifted 
space this symmetry corresponds to the invariance of the RD under a 
$\pi$-rotation around the center (or the reflection with respect to the center).
 From Fig.\ \ref{fig:fd}(a) it is seen that 
the image of zone 1 of strip 1 is in strip $+3$, hence still in 1 after
wrapping. On the contrary, the image $1'$ of strip 1 is in strip 2.
However, a $\pi$-rotation can put the image back into strip 1. So,
using the reflectional symmetry, we may focus only on strip 1, which
may be regarded as the minimal domain (MD). In this way the 5-letter symbolic
dynamics is reduced to a 3-letter one, and the $\pi$-rotation changes the
parity of $R_1$. More specifically speaking, we have the following ordering
for the minimal domain
\begin{equation}
\bullet L_0< \bullet R_0< \bullet R_1,\qquad L_0\bullet <R_0\bullet
<R_1\bullet ,
\label{md}
\end{equation}
and only $R_0$ in a leading string reverses the ordering. The symbolic
plane of the MD is shown in Fig.~\ref{fig:mdsp} where 34,000 points of several 
real orbits are drawn together with the pruning fronts. 
The primary pruning front of $\bullet D_0$ is roughly diagonal,
while that of $\bullet C_0$ is almost vertical. The latter encloses a
forbidden zone near the top in the symbolic plane, which is responsible
for many interesting bifurcations. In the 5-letter code for the RD, the second
and fourth quadrants of the symbolic plane are forbidden. The symbolic plane
for the 3-letter code of the MD looks more compact.

Let us discuss the pruning front in some detail. 
A point $Q\bullet D_0P$ on the partition line $\bullet D_0$ discards
a rectangle in the symbolic plane as a forbidden zone, the top-right corner
of which is the point $Q\bullet D_0P$, as shown in Fig.~\ref{fig:rd-orb}. 
When representing the point in the symbolic plane, we have replaced $D_0$ 
by its right limit, i.e., a point with $s=0_+$. Its allowed zone
is the rectangle, the lower-left corner of which is the same point. Similarly, a
point $Q\bullet C_0P$ discards a rectangle as its forbidden zone whose two
lower corners are $Q\bullet L_0P$ and $Q\bullet R_0P$. The point also
determines two allowed zones. The above mentioned two corners form the 
upper-right corner of the left allowed zone and the upper-left corner of the 
right one. We have numerically found the following 5 points on the 
partition lines:
\begin{eqnarray*}
 && P_1:\quad \cdots R_1L_0R_1L_0R_1L_0R_1^2R_0^2R_1^2 \bullet C_0R_1R_0R_1
		  R_0^2R_1^2R_0^2R_1 \cdots ,\cr
&&P_2:\quad \cdots R_1L_0R_1R_0R_1L_0R_0L_0R_1^2 \bullet C_0R_1R_0^2R_1^3
		  R_0R_1L_0R_1R_0R_1 \cdots ,\cr
&&P_3:\quad \cdots R_1^3L_0R_0^2R_1^2L_0R_1R_0R_1\bullet C_0R_1^2R_0^3
		  R_1^{12}L_0\cdots ,\cr
&&P_4:\quad \cdots R_1L_0R_1L_0R_1L_0R_1^4R_0R_1L_0R_1L_0R_0 \bullet L_0
		  R_1L_0R_1L_0R_1R_0R_1^4L_0R_1L_0R_1 \cdots ,\cr
&&P_5:\quad \cdots R_1L_0R_1L_0R_1^4R_0R_1L_0R_1L_0R_1L_0 \bullet R_0
	L_0R_1L_0R_1R_0R_1^4L_0R_1L_0R_1 \cdots ,\cr
\end{eqnarray*}
where the last two points are on $\bullet D_0$. The first tree points 
correspond to lower corners of the top three rectangles with dotted sides in 
Fig.~\ref{fig:rd-orb}, while the other two to the upper-right corners 
of the two bottom 
rectangles. It can be verified that the point $(R_0R_1^2R_0^3R_1L_0
R_1^2)^\infty\bullet (R_0R_1^2R_0^3R_1L_0R_1^2)^\infty$ is in the 
forbidden zone of $P_2$, so the periodic sequence $(R_0R_1^2R_0^3R_1
L_0R_1^2)^\infty$ is nonadmissible. One can verify the admissibility of 
the sequence $(R_1L_0R_1R_0R_1^2R_0R_1R_0R_1)^\infty$. In fact, its $0$, 
$1$, $2$, $4$, $5$, $7$, $9^{\rm th}$ shifts are in the allowed zone of $P_1$, while 
its $3$, $8^{\rm th}$ shifts in that of $P_3$. At the same time, its $0$, $1$, $3$, 
$5$, $6$, $8^{\rm th}$ shifts are in the allowed zone of $P_4$, while its $2$, $4$ 
$7$, $9^{\rm th}$ shifts in that of $P_3$. In this way we may examine the 
admissibility of periodic sequences from the known coding of partition 
lines. The above orbit $(R_1L_0R_1R_0R_1^2R_0R_1R_0R_1)^\infty$, written in the
5-letter code, is $(R_1R_2R_1R_0R_1^2R_0R_1L_2R_1)^\infty$. Symmetry of
sequences by means of the 5-letter code will be discussed in Sec. V.
  
\section{Correspondence between the axis codes and the bounce codes}
\label{sec:4disk}
In the literature the symbolic description based on bounces is used. Bounces 
are well defined for billiards. We may explore the relation between the axis 
codes and the bounce codes by examining a billiard. For simplicity, let us 
consider the touching four-disk billiard. As shown in Fig.\ \ref{fig:4d}(a), 
the boundary of 
the billiard is formed by the four disks which are labeled with a, b, c, and 
d, respectively. Taking the boundary as a Poincar\'e section, we have the 
corresponding Birkhoff coordinates similar to those of the stadium 
billiard, which we denote by $(r, u)$ to distinguish from the above 
$(s, v)$, and hence the associated map on the annulus \cite{Zheng1}. 
We normalize the boundary 
length as unity, and measure the length from the touching point of disks a 
and d. As shown in Fig.\ \ref{fig:4d}(b), the fundamental domain of the annulus consists of 
four strips corresponding to the four disks. The image of the strip of disk 
a, which has three pieces in three other strips, determines the partition of 
the reduced domain consisting of a single strip. The three regions of the RD 
are labeled with $\bullet 0$, $\bullet 1$, and $\bullet 2$ in the figure. By 
the argument based on lifting, we have the ordering
\begin{equation}
\bullet 0< \bullet 1 < \bullet 2.
\end{equation}
Since every bounce on a disk reverses the ordering of orbits, each symbol has an 
odd parity. That is, for forward symbolic sequences, a string of an odd length 
is odd, while that of an even length is even. We have similar ordering rules 
for backward symbolic sequences. This is our version of the symbolic dynamics 
described in Ref.\ \cite{Hansen1}.

When the strip of $r\in [0, 0.25)$ is divided at $r=0.125$, and an 
eighth of the FD is considered, the image of the line $r=0.125$ has three 
pieces respectively in strips b, c and d. The partition of a half strip, 
which is determined by putting all the pieces of its image together back into 
the half strip, will involve symbols more than three. 

To explore the correspondence between the axis codes and the bounce codes, we 
examine how an orbit starting from a point on an axis hits different disks. 
This gives the relation between the phase spaces $(s,v)$ associated with the 
axis contour and $(r,u)$ associated with the billiard boundary. Consider the 
orbit AC grazing against disk d in Fig.\ \ref{fig:cor}(a). It corresponds to a 
point of $s\in (-1,0)$, or equivalently a point of $s\in (1,2)$ when the RD 
is considered. Starting from point A on the 
$x$-axis, orbits right of this grazing orbit hit disk d before intersecting 
an axis again, while orbits on its left hit the $y$-axis directly before 
having a bounce on disk c. The orbit ABO, 
which passes through the origin, separates orbits crossing the $y$-axis from 
those crossing the $x$-axis. The orbits of the type AC correspond to line 
$\bullet C_2$ in the RD of the $s$-$v$ phase space since the line demarcates 
the order preserving region without a bounce from the order reversing region 
with a bounce. Similarly, the orbits of type ABO correspond to line $\bullet 
B_2$. Thus, bisecting the region $\bullet R_1$ into $\bullet R_+$ and 
$\bullet R_-$ with line $s=1$, we may re-label the right half of the RD as 
in Fig.\ \ref{fig:cor}(b). Comparing this figure with Fig.\ 3(b), we see 
that $\bullet R_+$ 
and $\bullet R_2$ belong to $\bullet 0$, which associates with orbits leaving 
from disk a and bouncing on disk d. The region $\bullet L_2$ has no 
direct correspondence with any bounce. Orbits from the region, after missing a 
bounce, will hit disk c before intersecting again an axis. We may re-label the 
region as $(\bullet 1)$. Consequently, we have the correspondence:
$$\bullet R_+, \bullet R_2\to \bullet 0; \quad \bullet L_2X\to \bullet 1,$$
where $X$ stands for any symbol. 

The other half of the RD in $s$-$v$ space may be analyzed in a similar way. 
The re-labeling of regions is also shown in Fig.\ \ref{fig:cor}(b), which 
indicates the correspondence:
$$\bullet R_-, \bullet R_0\to \bullet 2; \quad \bullet L_0X\to \bullet 1.$$

Putting Fig.\ \ref{fig:rd}(a) on the top of 3(b), we see that the allowed 
symbolic pair are
$$L_2R_+,L_2R_2;\ R_2R_+, R_2R_2, R_2L_2;\ R_-R_+, R_-R_2, R_-L_2;\quad
R_+R_-, R_+R_0, R_+L_0;\ R_0R_-, R_0R_0, R_0L_0;\ L_0R_-, L_0R_0.$$
Two rules can then be extracted:
\begin{enumerate}
\item $L_2$ and $L_0$ are always followed by a symbol of odd parity.
\item $R_+$ follows either $L_2$, or $R_2$, or $R_-$, while $R_-$ follows 
either $L_0$, or $R_0$, or $R_+$.
\end{enumerate}
The first rule explains that $L_0X$ or $L_2X$ has the same parity as the 
bounce code $1$. The second rule can be used to refine $R_1$ into $R_+$ and 
$R_-$, and then further to convert them into bounce codes. For example, an 
orbit shown in Fig.\ 9(a) has the axis codes $(R_2^2R_1R_0^2R_1)^\infty$. The 
codes are first refined as $(R_2^2R_+R_0^2R_-)^\infty$, and then converted to 
$(0^32^3)^\infty$. Another orbit shown in Fig.\ 9(b) has the codes 
$(R_2^2R_1^3L_0R_0R_1)^\infty$. This sequence is converted as follows:
$$(R_2^2R_1^3L_0R_0R_1)^\infty\to (R_2^2R_+R_-R_+L_0R_0R_-)^\infty\to 
(0^32012)^\infty .$$

The ordering (\ref{ford}) can be refined as 
$$\bullet L_0< \bullet R_0< \bullet R_-<\bullet R_+ < \bullet R_2< 
\bullet L_2, $$
which, under a cyclic transformation, becomes
$$\bullet R_+ < \bullet R_2< \bullet L_2< \bullet L_0< \bullet R_0< 
\bullet R_-.$$
This ordering is then converted to 
$$\bullet 0 <\bullet 1< \bullet 2,$$
which is just the ordering of the bounce codes. A detailed analysis of the 
relation between axis codes and bounce codes will be presented elsewhere.

\section{Symbolic sequences and symmetry of orbits}
\label{sec:encd_class}
The configurational space of the original dynamics is the half $\rho$-$z$ 
plane, which converts to the first quadrant of the $\mu$-$\nu$ plane. The 
rotational symmetry in the $\mu$-$\nu$ plane is the invariance under the 
transformation from $(\mu ,\nu ,p_\mu ,p_\nu)$ to $(-\nu ,\mu ,-p_\nu 
,p_\mu)$, $(-\mu ,-\nu ,-p_\mu ,-p_\nu)$ and $(\nu ,-\mu ,p_\nu ,-p_\mu)$ 
(denoted by $\rho$, $\pi$ and $\bar\rho$). This symmetry has nothing to do 
with the original dynamics. Confining the dynamics to the RD, we remove the 
symmetry. 

As for the reflectional symmetry, we may focus only on 
$\sigma_\mu :\ (\mu ,-\nu ,p_\mu ,-p_\nu)$ since we can composite 
$\sigma_\nu =\pi\circ\sigma_\mu$, $\sigma_{13}=\sigma_\mu\circ\bar\rho$ and 
$\sigma_{24}=\sigma_\mu\circ\rho$. We shall denote $\sigma_\mu$ simply by 
$\sigma$. While $\rho$, $\pi$ and $\bar\rho$ have 
no effect in the $s$-$v$ space, $\sigma$ leads to $(s,v)\to (2-s,-v)$, 
which, written in symbols, is $L_0 \leftrightarrow L_2$, 
$R_0 \leftrightarrow R_2$ and $R_1 \leftrightarrow R_1$. To discuss the 
reflectional symmetry, the 5-letter code must be used. Thus, if an orbit 
is symmetric under reflection, its symbolic sequence is invariant up to a 
shift under the symbolic transformation of $\sigma$. For example, 
sequence $(R_1R_2^2R_1R_0^2)^\infty$ of period 6 is symmetric under 
$\sigma$, while $(R_1^2R_0^4)^\infty$ is not. This two orbits are shown 
in Fig.~9(a).

The Hamiltonian dynamics has the time-reversal symmetry: $(\mu, \nu ,p_\mu , 
p_\nu ;\tau )\to (\mu, \nu ,p_\mu ,p_\nu ;-\tau )=(\mu, \nu ,-p_\mu ,
-p_\nu ;\tau )$. This transformation, denoted by $T$, leads to $(s,v)\to 
(2-s, v)$ in the $s$-$v$ space. Written in symbols, this means 
$\bullet L_0\leftrightarrow L_2\bullet$, $\bullet R_0\leftrightarrow 
R_2\bullet$, $\bullet R_1\leftrightarrow R_1\bullet$, $\bullet R_2
\leftrightarrow R_0\bullet$ and $\bullet L_2\leftrightarrow L_0\bullet$. 
Thus, the effect of $T$ on a sequence is to reverse the sequence and 
apply $\sigma$ to it. For example, transformation $T$ converts 
sequence $(R_1^3L_0R_0R_1R_2^2)^\infty$ to $(R_0^2R_1R_2L_2R_1^3)^\infty$.
 We may combine $T$ and $\sigma$ to 
define the transformation $T_\sigma$ of 
reversal by $T_\sigma =T\circ\sigma =\sigma\circ T$. 
In the 3-letter symbolic dynamics we identify $L_2$ with $L_0$, and $R_2$ 
with $R_0$, so $T$ just reverses a sequence. Transformation $T_\sigma$ converts
the above just mentioned orbit $(R_1^3L_0R_0R_1R_2^2)^\infty$ to 
$(R_2^2R_1R_0L_0R_1^3)^\infty$. The two orbits $(R_1^3L_0R_0R_1R_2^2)^\infty$
and $(R_2^2R_1R_0L_0R_1^3)^\infty$ are shown in Fig.~9(b). (In the configuration
space orbits $(R_1^3L_0R_0R_1R_2^2)^\infty$ and $(R_0^2R_1R_2L_2R_1^3)^\infty$
are not distinguishable.)

The transformations $\sigma$, $T$ and $T_\sigma$ may be used to generate 
admissible sequences from a known one. When performing a transformation on 
a sequence results in the same sequence up to a shift, the sequence is 
symmetric. In Table I, we give the non-repeating strings of 15 admissible 
periodic sequences, the initial values $(s, v)$ and the symmetry of their 
corresponding orbits. 

Orbits which pass through the origin play a special role in the 
semiclassical theory of closed orbits. Let us analyze their symmetry in some 
detail. Right at the origin the momentum $v$ is ill-defined. However, for an 
orbit passing through the origin we may consider its left and right limits 
since both of them are well-defined. It can be seen that around the origin 
we have codes $BD_0^2$ and $BD_2^2$ for the two limits, where $B$ stands 
for $B_0$ or $B_2$. In the RD $D_0$ should have been identified with $D_2$ 
since the dynamics is on the annulus. Here by $D_0$ and $D_2$ we mean 
$s=0_+$ and $2_-$, respectively. It can be further verified that the first
$D_0(D_2)$ and second $D_2(D_0)$ are related as $(s,v)\leftrightarrow 
(2-s,v)$, which is just the transformation $T$. This means that, after 
identifying $D_0$ with $D_2$, the sequence of an orbit passing through the 
origin must be $T$-invariant. For example, the orbit shown in Fig.~9(c) 
has the sequences $(B_2D_2^2R_0B_0D_0^2R_2)^\infty$ and $(B_0D_2^2R_2
B_2D_0^2R_0)^\infty$ as its right and left limits. Written more precisely, 
the two limiting sequences are $(R_2L_2R_1R_0R_0L_0R_1R_2)^\infty$ and 
$(R_2R_1L_0R_0R_0R_1L_2R_2)^\infty$, respectively. The $T$ transformation 
of the former is $(R_0R_1L_2R_2R_2R_1L_0R_0)^\infty$, which turns out to be 
just the 4-th shift of the latter. (A further observation discovers that the 
codes $BDD\cdots$ can be equally well written as $BDB\cdots$ since the third 
point is on the common segment of $\bullet D$ and $\bullet B$, where $D$, 
similar to $B$, stands for $D_0$ or $D_2$.)

Many short closed orbits have been found in Ref.~\cite{DD}. Non-repeating 
strings or their doubled strings of the periodic sequences corresponding 
to these orbits are listed 
in Table II, where sequences and initial values $v$ are for the right limit 
orbits starting at $s=0^+$. Sequences for left limit orbits may be obtained by exchanging 
$D_0$ with $D_2$. While the labeling of Ref.~\cite{DD} for orbits are given 
in the table, we arrange them according to the order of sequences and 
symmetry, and shall refer this numbering to orbits. 
More precisely speaking, the two sequences which relate to each 
other by $\sigma$ are always listed successively, and ordered according to 
one of the two without referring to the order of the other. The orbit shown 
in Fig.~9(c) is orbit 27. 

From the table 
it is clearly seen that symbolic coding provides a way to classify orbits. 
For example, according to the coding orbits $1,3,5,15,25$ and $2,4,10,20$ 
form a big family of the type $BD^2R_1^{n}BD^2R_1^{n}$. When $n$ is odd the 
non-repeating string of an orbit reduces to half. We have doubled it in the 
table to make it look more like that for an even $n$. There are two more 
orbits (29 and 30) for which doubled strings are given. 
We can see many pairs of orbits which should be created in a 
bifurcation related to $\bullet C$. A 
pair of orbits 29 and 30 are given in Fig.~9(d) to show that they have the 
same topology. Other examples are orbits 6 and 8, orbits 11 and 13, and 
orbits 28 and 30. In the table 8 orbits are obtained by the transformation 
$\sigma$, which is indicated in the last column of the table.
The initial values of $v$ for two such orbits related by $\sigma$ 
are different only in their sign, so only one $v$ is given. 
The sequences listed for these orbits then correspond to a left limit.
\section{Conclusion}
\label{sec:sum}

In the above we have given a natural way to construct symbolic dynamics for
the the diamagnetic Kepler problem. Our choice of the Poincar\'e section
avoids any ambiguity in identifying bounces. A lift of the phase space
helps us to understand the dynamics. Tangencies between stable and unstable
manifold foliations play an important role in the construction of symbolic
dynamics. Based on the symbolic coding we have analyzed the rotational and
reflectional symmetry, and the time reversal symmetry as well. Symbolic
coding provides us a convenient way to classify orbits. For a billiard limit 
we have established the relation between our coding and the coding based on 
bounces. 

\acknowledgments
{ This work was supported in part by the Post-Doctoral Foundation of
China (ZBW) and the National Natural Science Foundation of China (WMZ).}



\begin{table}[htbp]
\label{tab1}
\begin{footnotesize}
Table I. Some orbits which do not pass through the origin.
\vspace{0.5cm}
\begin{tabular}{rcclrcl}
No.& $s$ &$v$    &Sequence &Period & Symmetry &Bound Codes\\
\tableline
1 &0.502590559992&0.000000000377& $R_0R_1R_2^2R_1R_0$       &6 &$\sigma, T, T_\sigma$  	&$2^20^32$\\
2 &0.525466094731&0.083479979410& $R_0^2L_0R_1^2R_0$        &6 &  			&$2^2102$\\
3 &0.597726644479&0.000000000105& $R_0$                     &4 & $T_\sigma$ 		&$2^4$\\
4 &0.627764854085&0.000000000692& $R_0^2R_1^2R_0^2$         &6 &$T_\sigma$ 		&$2^302^2$\\
5 &0.634615582851&0.343668565696& $R_1L_2R_1R_0$            &4 &$T_\sigma$  		&$212$\\
6 &0.634764050232&0.000000004571& $R_0^2R_1^4R_0^2$         &8 &$T_\sigma$  		&$2^30202^2$\\
7 &0.637932489171&0.466976827386& $R_1R_2R_1R_0$            &4 &$\sigma, T, T_\sigma$ 	&$20^22$\\
8 &0.644118560693&0.000000282982& $R_0^2R_1^{12}R_0^2$      &16 &$T_\sigma$ 		&$2^3(02)^62$\\
9 &0.645908548293&0.860542468032& $R_1^7R_2^2L_2R_1^3L_0R_0L_0$ &16 & 			&$1(02)^30^21201$\\
10&0.653504728307&0.022526445642& $R_0L_0R_1^{22}R_0^2$     &26  & 			&$21(02)^{11}2$\\
11&0.653578168108&0.014059036967& $R_0L_0R_1^{26}R_0^2$     &30 & 			&$21(02)^{13}2$\\
12&0.671107671805&0.000000002635& $R_0L_0R_1^4L_0R_0$       &8 &$T_\sigma$  		&$210201$\\
13&0.685136083296&0.257524880819& $R_1R_2R_1R_0^2L_0R_1^8R_0^2$ &16 & 			&$20^22^210(20)^32^2$\\
14&0.714849526728&0.061364237283& $R_1R_2R_1R_0R_1^2R_0R_1L_2R_1$ &10 & 		&$20^22^202^21$\\
15&0.744712436526&0.064776613634& $R_1^2L_0R_0R_1R_2^2R_1$  &8  &  			&$20120^3$\\
\end{tabular}
\end{footnotesize}
\end{table}

\begin{table}[htbp]
\label{tab2}
\begin{footnotesize}
Table II. Sequences of orbits passing through the origin. By $O_i$ we mean
the $i$-th orbit according to the numbering in the first column.
\vspace{0.5cm}
\begin{tabular}{rrclrcl}
No.& label$^{[2]}$& $v$  &Sequence & Period& Symmetry &Bound Codes\\
\tableline
1& 30  & 0.974971232297&$D_0R_1^8B_2D_0^2 R_1^8B_0D_0$ &22 &		&$21(02)^401(20)^4$\\
2& 24  &0.972898880448 &$D_0R_1^7B_0D_2^2 R_1^7B_0D_0$ &20 &		&$21(02)^410(20)^3$\\
3& 19  &0.970340776638 &$D_0R_1^6B_2D_0^2 R_1^6B_0D_0$ &18 &		&$21(02)^301(20)^3$\\
4& 13  &0.967086416031 &$D_0R_1^5B_0D_2^2 R_1^5B_0D_0$ &16 &		&$21(02)^310(20)^2$\\
5& 9   &0.962773942237 &$D_0R_1^4B_2D_0^2 R_1^4B_0D_0$ &14 &		&$21(02)^201(20)^2$\\
6& 29  &0.959653990845 &$D_0R_1^4R_2B_2D_2^2 R_0R_1^4B_0D_0$ &16 &	&$21(02)^20^212^2020$\\
7& 14  &     	       &$D_2R_1^4R_0B_0D_0^2 R_2R_1^4B_2D_2$ &16 &$\sigma\circ O_{6}$ 	&$2^210^2(20)^21(20)^2$\\
8& 28  &0.958697661319 &$D_0R_1^4L_2B_2D_0^2 L_0R_1^4B_0D_0$ &16 & 			&$21(02)^21^3020$\\
9& 15  &     	       &$D_2R_1^4L_0B_0D_2^2 L_2R_1^4B_2D_2$ &16 &$\sigma\circ O_{8}$ 	&$201(20)^21^320$\\ 
10& 8  &0.956722093284 &$D_0R_1^3B_0D_2^2 R_1^3B_0D_0$         &12 & 			&$21(02)^21020$\\
11& 27 &0.952489366877 &$D_0R_1^3R_0B_0D_0^2R_2R_1^3B_0D_0$  &14 & 			&$21(02)^2210^220$\\
12& 23 &  	       &$D_2R_1^3R_2B_2D_2^2R_0R_1^3B_2D_2$  &14 &$\sigma\circ O_{11}$	&$12(20)^21(20)^20$\\
13& 20 &0.950042745974 &$D_0R_1^3L_0B_0D_2^2 L_2R_1^3B_0D_0$       &14 & 		&$210201^320$\\
14& 22  &   &$D_2R_1^3L_2B_2D_0^2 L_0R_1^3B_2D_2$ &14 &$\sigma\circ O_{13}$		&$1^202012021$\\
15& 5   &0.947457897321 &$D_0R_1^2B_2D_0^2R_1^2B_0D_0$         &10& 			&$21020120$\\
16& 17  &0.940875954726 &$D_0R_1^2R_2B_2D_2^2 R_0R_1^2B_0D_0$ &12& 			&$21020^212^20$\\
17& 18  &    &$D_2R_1^2R_0B_0D_0^2 R_2R_1^2B_2D_2$ &12 &$\sigma\circ O_{16}$		&$2^210^220120$\\
18& 12  &0.935006464896 &$D_0R_1^2L_2B_2D_0^2 L_0R_1^2B_0D_0$       &12& 		&$21021^30$\\
19& 16  &     &$D_2R_1^2L_0B_0D_2^2 L_2R_1^2B_2D_2$ &12 &$\sigma\circ O_{18}$		&$201^3201$\\
20& 3   &0.931024837844 &$D_0R_1B_0D_2^2R_1B_0D_0$         &8& 				&$210210$\\
21& 11  &0.918172430278 &$D_0R_1R_0B_0D_0^2R_2R_1B_0D_0$       &10 & 			&$2102^210^2$\\
22& 10  &    &$D_2R_1R_2B_2D_2^2R_0R_1B_2D_2$&10 &$\sigma\circ O_{21}$			&$12^20120^2$\\
23& 7 &0.899732522729 &$D_0R_1L_0B_0D_2^2 L_2R_1B_0D_0$       &10 & 			&$2101^3$\\
24& 6 &  &$D_2R_1L_2B_2D_0^2 L_0R_1B_2D_2$  &10 &$\sigma\circ O_{23}$			&$1^20121$\\
25& 2   &0.891672484472 &$D_0B_2D_0^2B_0D_0$       &6 & 				&$2101$\\
26& 21  &0.864842403049 &$D_0R_2^2B_2D_0^2R_0^2B_0D_0$ &10  &	 			&$210^312^2$\\
27& 4   &0.849666377360 &$D_0R_2B_2D_2^2R_0B_0D_0$         &8  & 			&$210^212$\\
28& 26  &0.804140836919 &$D_0R_2R_1^5L_0B_0D_0^2L_2R_1^5R_0B_0D_0$ &20  &		&$1^3(20)^22^210^2(20)^2$\\
29& 25  &0.803121053916 &$D_0R_2R_1^5R_0B_0D_2^2R_2R_1^5R_0B_0D_0$ &20  &		&$210(02)^3210(02)^3$\\
30& 31  &0.790624022578 &$D_0L_2R_1^5L_0B_0D_2^2L_2R_1^5L_0B_0D_0$ &20  &		&$1^3(20)^21^3(20)^2$\\
31& 1   &0.707106780832 &$D_0D_2^2D_0$             &4 &					&$1^2$\\
\end{tabular}
\end{footnotesize}
\end{table}

\begin{figure}
\caption{Boundary of the transformed potential for the diamagnetic 
Kepler problem at $\epsilon=0$, and a typical orbit.}
 \label{fig:orb1}
\end{figure}

\begin{figure}
\caption{The image (a) and preimage (b) of the fundamental domain.}
 \label{fig:fd}
\end{figure}

\begin{figure}
\caption{The partition of the reduced domain according to preimage (a)
and to present (b).}
 \label{fig:rd}
\end{figure}

\begin{figure}
\caption{(a) Symbolic plane of the reduced domain. Approximately 
68000 real orbit points are shown.
(b) Primary pruning front. Points on $\bullet D_0$ and $\bullet D_2$ are 
marked with diamonds, and points on $\bullet C_0$ and $\bullet C_2$ with 
crosses. }
\label{fig:rdsp}
\end{figure}

\begin{figure}
\caption{Symbolic plane of the minimal plane. Approximately 34000 
real orbit points are shown together with the primary pruning front. 
Points on $\bullet D_0$ are marked with diamonds, and points 
on $\bullet C_0$ with crosses.  }
\label{fig:mdsp}
\end{figure}

\begin{figure}
\caption{Admissible and forbidden orbits are determined by five points 
on two partition lines. }
\label{fig:rd-orb}
\end{figure}

\begin{figure}
\caption{The touching four-disk billiard (a) and its phase space (b). 
Here the billiard boundary is taken as a Poincar\'e section. The Birkhoff 
coordinates are denoted by $(r,u)$.}
\label{fig:4d}
\end{figure}

\begin{figure}
\caption{The grazing orbit and center-passing orbit (a) determine 
the correspondence between $(s,v)$ space and the bounce-based $(r,u)$ space 
(b).}
\label{fig:cor}
\end{figure}

\begin{figure}
\caption{Examples of orbits: (a) $\sigma$-symmetric 
$(R_1R_2^2R_1R_0^2)^\infty$ (solid) and asymmetric $(R_1^2R_0^4)^\infty$ (dashed); (b) an orbit 
$(R_1^3L_0R_0R_1R_2^2)^\infty$ (solid) and its $T_{\sigma}$-transformation image 
$(R_2^2R_1R_2L_2R_1^3)^\infty$ (dashed); (c) an orbit passing through the origin 
(orbit 27 in Table II); (d) orbits $D_0R_2R_1^5L_0B_0D_0^2L_2R_1^5R_0B_0D_0$ (solid)
and $D_0R_2R_1^5R_0B_2D_2^2R_2R_1^5R_0B_0D_0$ (dashed) with the same topology.}
\label{fig9}
\end{figure}

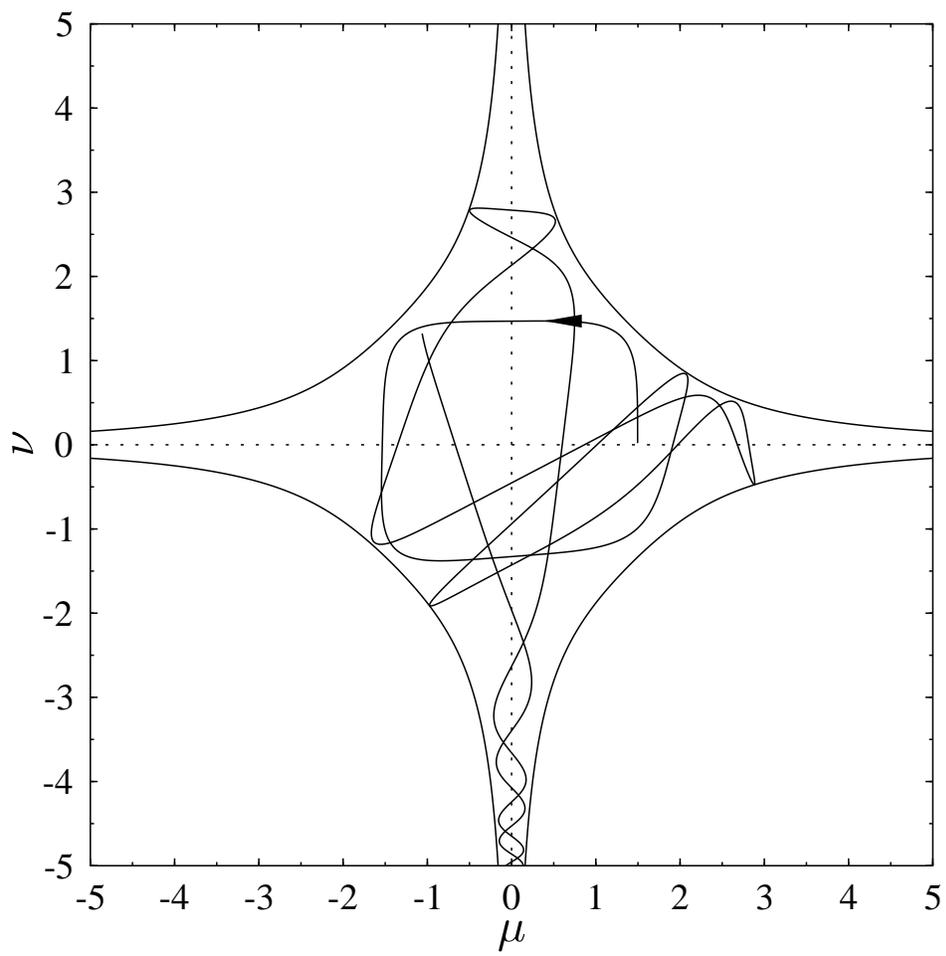
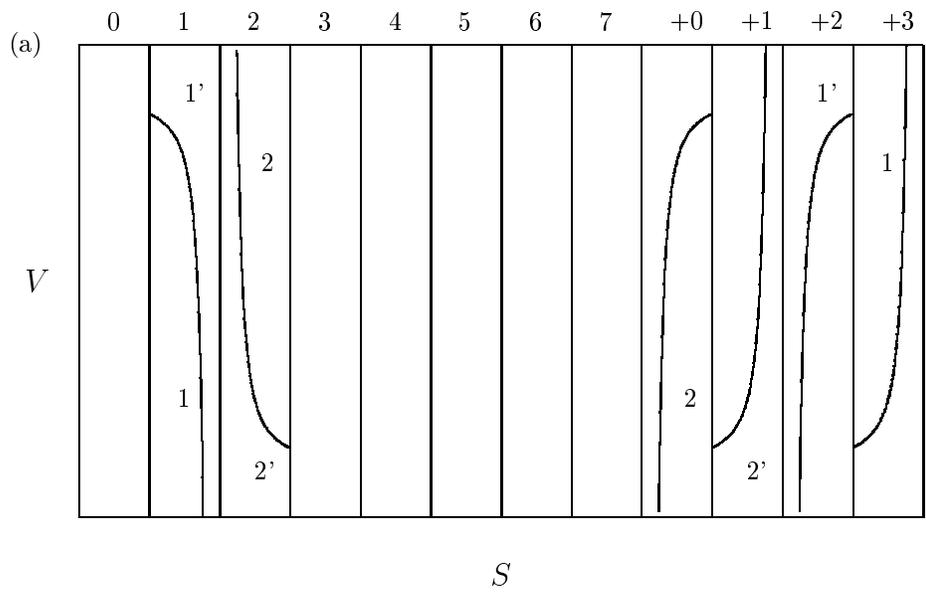
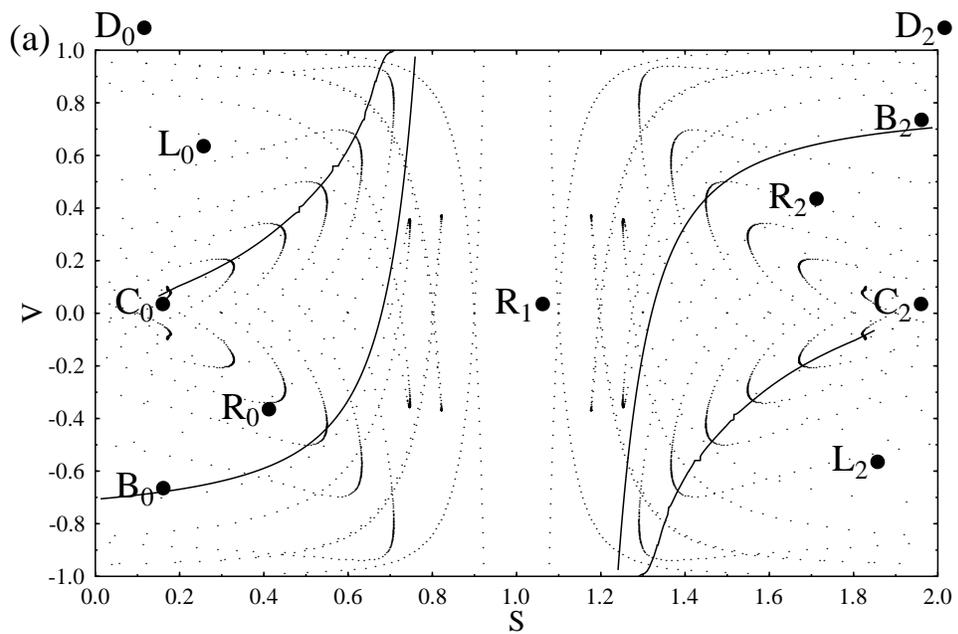
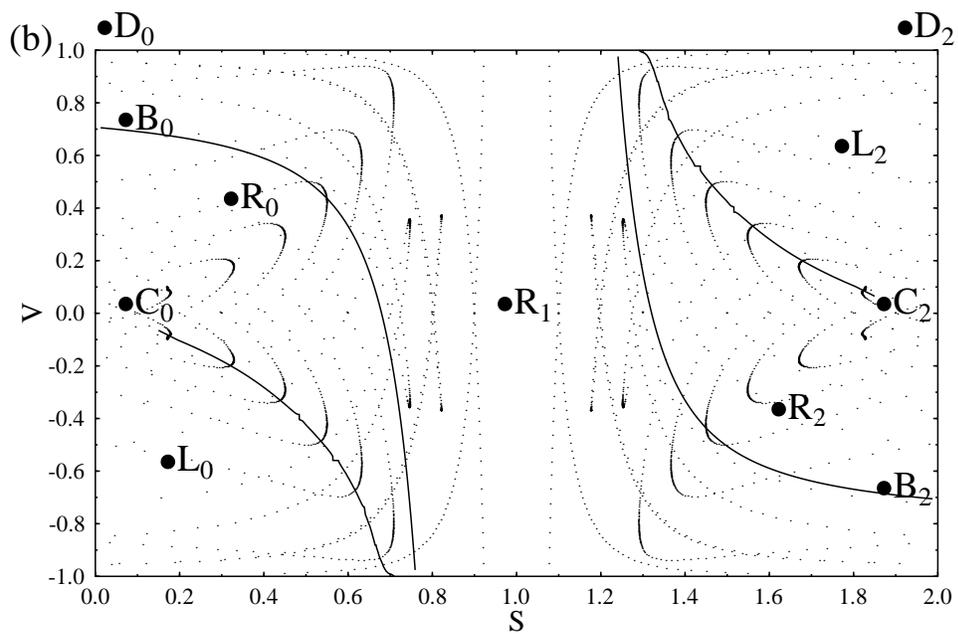
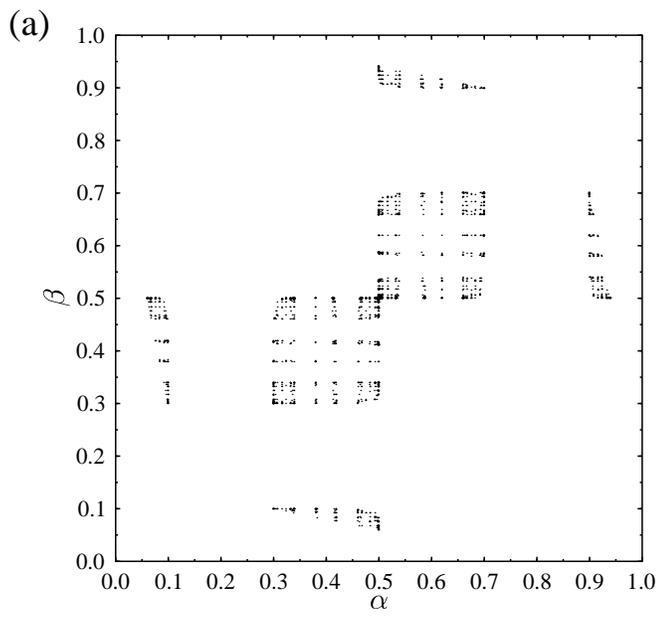
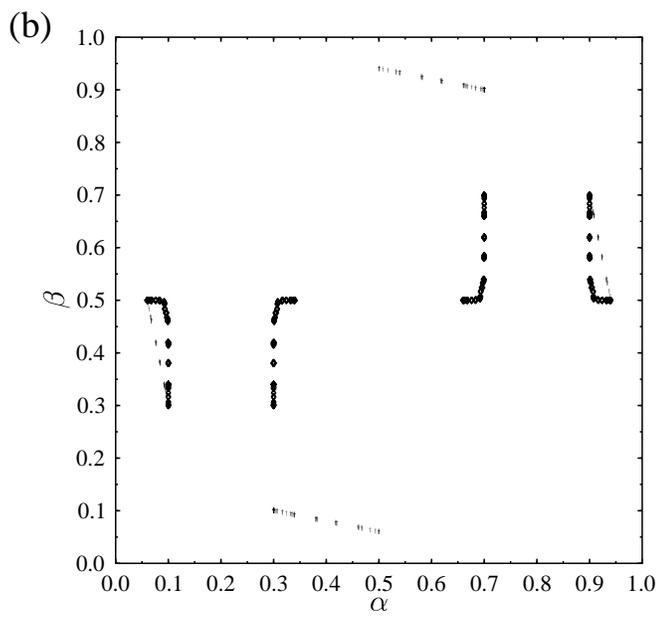
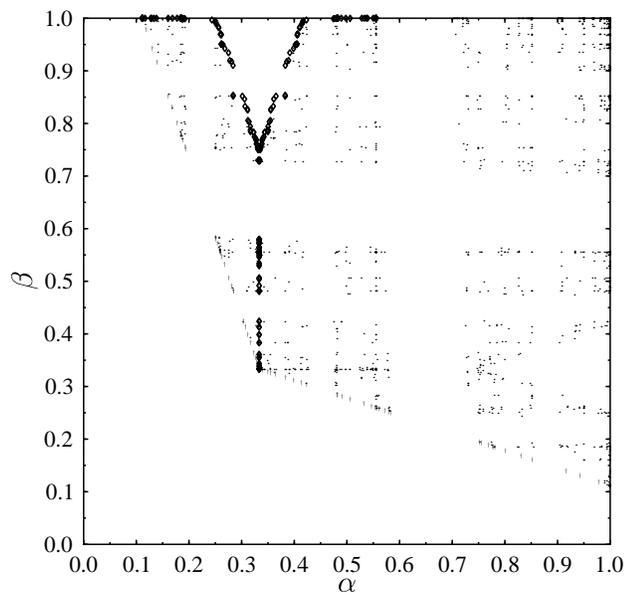
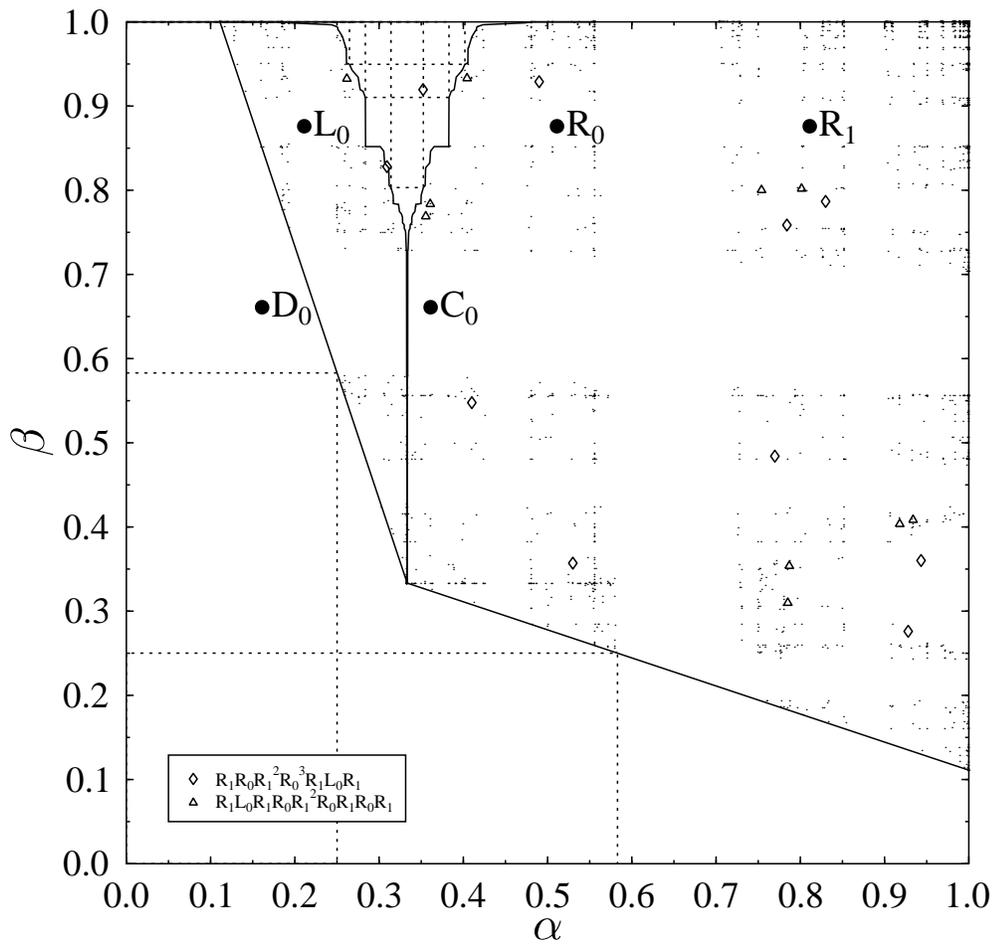
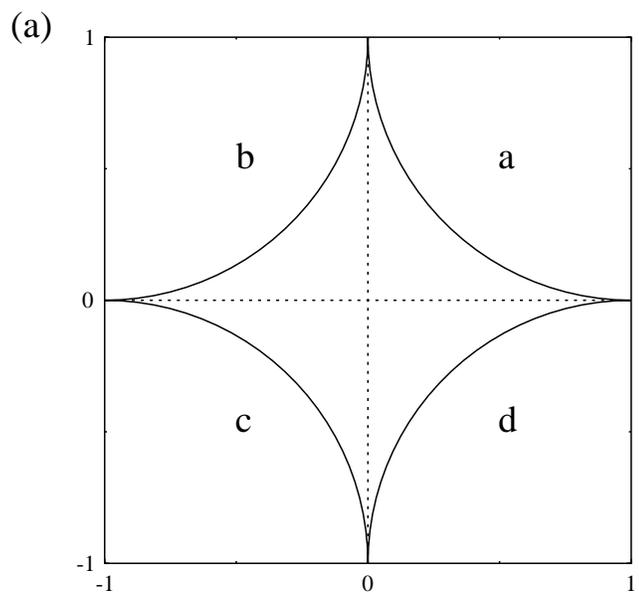
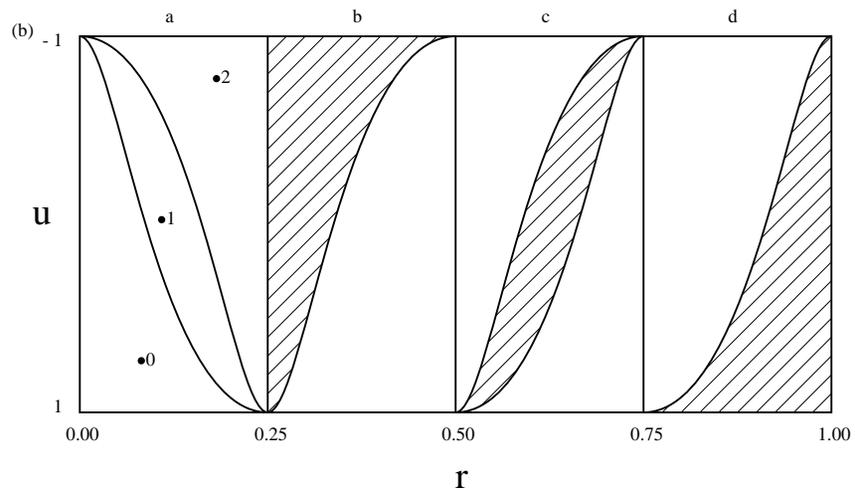
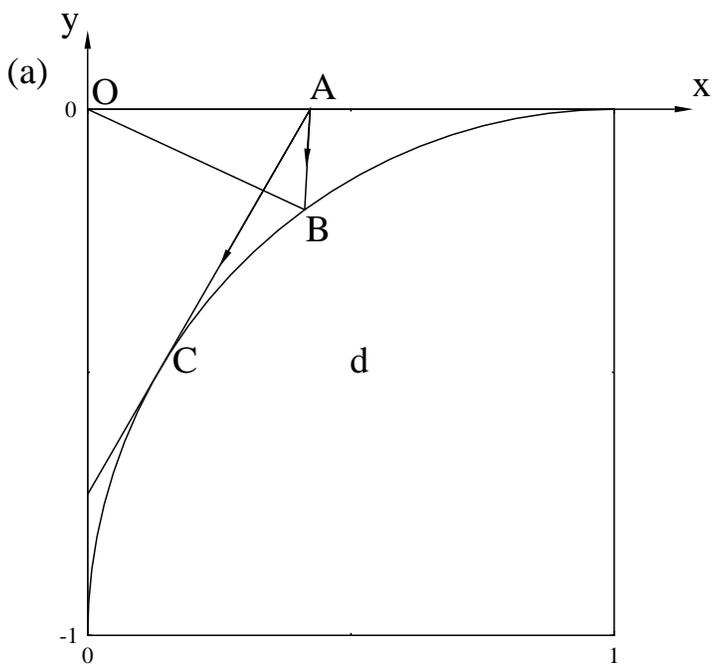
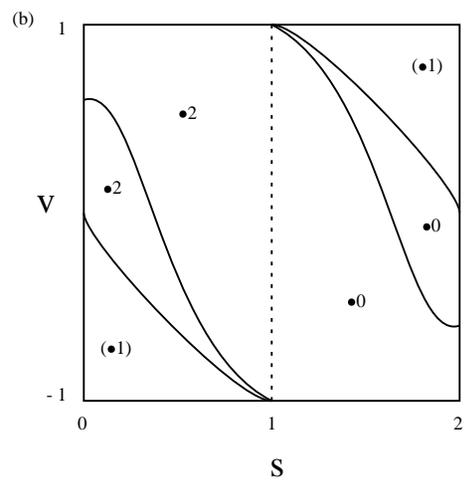
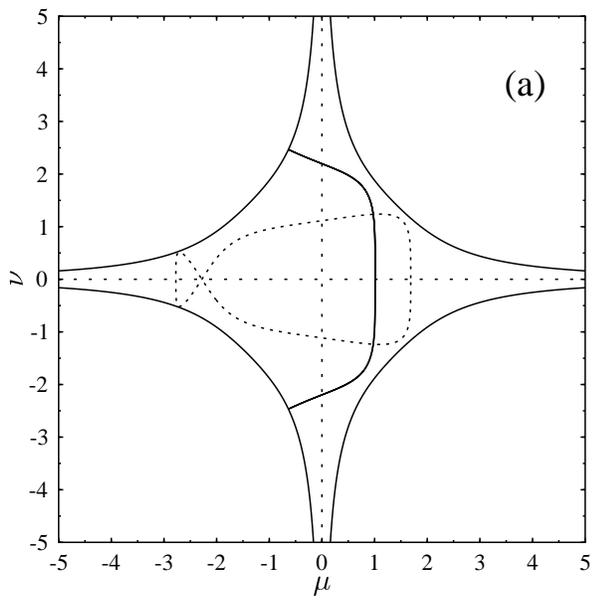
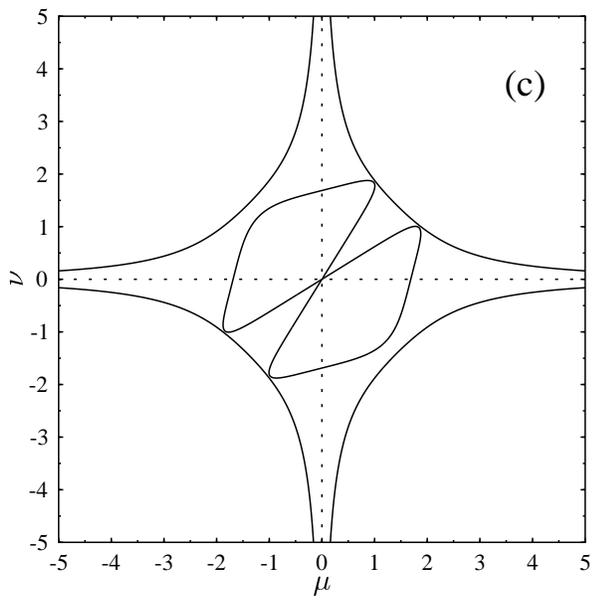
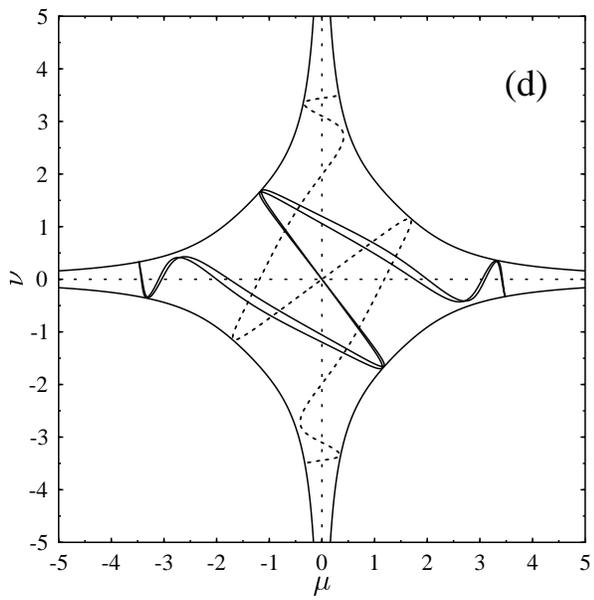
\end{document}